\documentclass[
	software,
	article,
	submit,
	pdftex,
	moreauthors
]{Definitions/mdpi} 

\renewcommand{\linenumbers}{}

\firstpage{1} 
\pubvolume{1}
\issuenum{1}
\articlenumber{0}
\pubyear{2026}
\copyrightyear{2026}
\datereceived{ } 
\daterevised{ } 
\dateaccepted{ } 
\datepublished{ } 

\Title{Towards a Software Architecture Description for Tax Compliance}

\Author{
Michael~Dorner~$^{1,}$*\orcidA{}, 
Oliver~Treidler~$^{2}$\orcidB{}, 
Tom-Eric~Kunz~$^{2}$\orcidC{}, 
Ehsan~Zabardast~$^{3}$\orcidD{}, 
Daniel~Mendez~$^{3,4}$\orcidE{}, 
Darja~Šmite~$^{3}$\orcidF{}, 
Maximilian~Capraro~$^{2}$\orcidG{} 
and Krzysztof~Wnuk~$^{3}$\orcidH{}
}

\AuthorNames{Michael Dorner, Oliver Treidler, Tom-Eric Kunz, Ehsan Zabardast, Daniel Mendez, Darja Šmite, Maximilian Capraro and Krzysztof Wnuk}

\address{%
$^{1}$ Technische Hochschule Nürnberg Georg Simon Ohm, Germany\\
$^{2}$ Kolabri GmbH, Germany\\
$^{3}$ Blekinge Institute of Technology, Sweden\\
$^{4}$ fortiss, Germany
}

\corres{Correspondence: michael.dorner@th-nuernberg.de}

\abstract{%
The internal reuse of software components owned by organizational units in different countries constitutes an implicit form of licensing that may be taxable under international regulations, making tax authorities an often overlooked stakeholder in software architecture.
This study assesses to what extent a deliberately minimal software architecture description can make implicit cross-border licensing visible and interpretable for tax experts.
We define a deliberately minimal architecture viewpoint based on components, dependencies, ownership, and jurisdiction, and construct a view for a large-scale industrial microservice software system comprising 2{,}518 components and 16{,}533 dependencies. The resulting architecture description is evaluated in a judgment study with four experienced tax auditors and tax advisors using semi-structured interviews.
The results show that the software architecture description provides a structured, evidence-based starting point for identifying cross-border reuse and supports tax audit discussions. However, experts consistently report limitations, including mismatches between software engineering and legal notions of ownership, unclear jurisdictional assignments, and the perceived risk of interpreting dependency counts as indicators of economic value. These limitations persist despite detailed architectural data. 
We conclude that software architecture descriptions can expose structurally relevant reuse relationships but cannot, on their own, support the legally meaningful interpretations required for tax assessment, indicating a fundamental mismatch between architectural abstractions and taxation concepts.
}

\keyword{software architecture; architecture viewpoints; global software engineering; microservices; taxation; transfer pricing; dependency analysis; cross-border software reuse; tax compliance}

\usepackage[T1]{fontenc}

\usepackage[autostyle]{csquotes}
\newcommand*{\myenquote}[1]{\enquote{\textit{#1}}}

\usepackage{graphicx}
\usepackage{textcomp}
\usepackage{xcolor}
\usepackage{fontawesome}
\usepackage{worldflags}
\usepackage{booktabs}
\usepackage{tabularx}
\usepackage{caption}

\usepackage{multirow}

\newcolumntype{L}[1]{>{\raggedright\arraybackslash}p{#1}}

\usepackage{tikz}
\usetikzlibrary{automata, positioning, arrows, arrows.meta, decorations.pathreplacing, calc, chains, shapes.misc, backgrounds, fit, perspective, matrix, shapes, shadows.blur}

\DeclareRobustCommand{\softwarecomponenttikz}{%
  \tikz[baseline=-0.5ex]{\node[rectangle, draw, inner sep=2mm, outer sep=0pt, fill=white]{};}%
}

\DeclareRobustCommand{\dependencytikz}{%
  \tikz[baseline=-0.5ex]{\draw[ACMRed, thick, -stealth] (0,0) -- (0.5,0);}%
}

\definecolor{ACMYellow}{RGB}{255, 214, 0}
\definecolor{ACMOrange}{RGB}{252, 146, 0}
\definecolor{ACMRed}{RGB}{253, 27, 20}
\definecolor{ACMLightBlue}{RGB}{131, 206, 226}
\definecolor{ACMGreen}{RGB}{166, 188, 9}
\definecolor{ACMPurple}{RGB}{101, 1, 107}
\definecolor{ACMDarkBlue}{RGB}{9, 53, 122}

\begin{document}

\section{Introduction}

Code is not tied to a physical, geographical location due to its intangible nature. The owner of code, however, is. The geographical location of those code owners has a legal implication that is often underestimated: Internally using software with owners in different countries is taxable, as it constitutes---from a taxation perspective---an often implicit form of cross-border licensing \citep{Dorner2022taxing,OECD2022}. To comply with international taxation standards \citep{OECD2022}, multinational enterprises must therefore be able to explain and justify the implicit cross-border licensing of their software systems. This makes tax authorities stakeholders in software architecture, as they have distinct concerns regarding cross-border software reuse. These concerns create a need for software architecture descriptions that systematically expose software structure and reuse across organizational and jurisdictional boundaries.

However, there is currently no established software architecture viewpoint that explicitly addresses the concerns of tax authorities and defines conventions for constructing corresponding architecture views for describing the implicit cross-border licensing of their software systems.

The objective of this study is to assess to what extent a deliberately minimal software architecture description can make implicit cross-border licensing visible and interpretable for tax experts.

Accordingly, we address the following research question: \emph{To what extent can a deliberately minimal software architecture description make implicit cross-border licensing in multinational enterprises visible and interpretable for tax experts?}

To this end, we define a deliberately minimal software architecture viewpoint grounded in software engineering understandings of taxation concerns as reflected in prior research and professional practice. The viewpoint relies exclusively on software components, dependencies between components, their owners, and the jurisdictions of those owners. We construct an architecture view in accordance with this viewpoint for a large-scale microservice architecture of a multinational enterprise, yielding a software architecture description that we evaluate with four experienced tax auditors and tax advisors in a judgment study based on semi-structured interviews. \Cref{fig:study_overview} provides an overview of the study. The purpose of this evaluation is not to propose an operational solution for tax compliance, but to empirically assess where software architecture abstractions support tax-relevant interpretations and where they systematically break down.

\begin{figure*}
	\begin{tikzpicture}[
    x=3cm,
    y=2cm,
    every node/.style={font=\small},
    activity/.style={draw, font=\small, text centered, minimum height=2em, fill=white},
    entry/.style={draw, circle, inner sep=1mm, fill=black},
    every label/.style={font=\small},
    arrowlabel/.style={midway, above=2mm, font=\small\bfseries, anchor=base},
    section/.style={fill=white, draw, blur shadow={shadow opacity=25, shadow xshift=1mm, shadow yshift=-1mm}, inner sep=3mm, rounded corners, }
  ]

  \node[entry] (concern) at (-2.1,1) {};

  \node[minimum height=2em, %
    label={[name=startlabel]below:\faFileTextO{} \Cref{sec:concernstakeholder}}
  ] (start) at (concern) {};

  \node[activity, label={[name=definelabel]below:\faFileTextO{} \Cref{sec:viewpoint}}] (define) at (-1,1) {Define viewpoint};

  \node[activity, label={[name=constructlabel]below:\faFileTextO{} \Cref{sec:view}}] (construct) at (0,0) {Instantiate view};

  \node[activity, text width=2.8cm, label={below:\faFileTextO{} \Cref{sec:result_viewpoint}}] (evaluate_viewpoint) at (1.5,1) {Evaluate viewpoint};
  
  \node[activity, text width=2.8cm, label={[name=evaluate_view_label, font=\small]below:\faFileTextO{} \Cref{sec:result_view}}] (evaluate_view) at (1.5,0) {Evaluate view};

  \node[entry, yshift=-0mm] (sa) at (-2.1,0) {};

  \draw[thick, -latex] (start) -- (define) node[arrowlabel] {Concern} ;
  \draw[thick, -latex, rounded corners] (define) -| ([yshift=0mm]construct);
  \draw[thick, -latex] (construct) -- (evaluate_view) node[near start, above, arrowlabel] (viewlabel) {View};
  \draw[thick, -latex] (define) -- (evaluate_viewpoint) node[arrowlabel, pos=0.35] {Viewpoint};
  \draw[thick, -latex] (sa) -- (sa-|construct.west) node[arrowlabel] {Large-scale software system};

  \begin{scope}[on background layer]
    \node[section, fit=(evaluate_view_label)(evaluate_viewpoint), label distance=4mm, label={[name=evaluate_view_label, font=\small, label distance=2mm]below:\faFileTextO{} \Cref{sec:results}}, label={[text width=4cm, text centered, font=\itshape\small, label distance=2mm]above:Evaluation with tax experts}] (section5) {};
  \end{scope}

  \begin{scope}[on background layer]
    \node[section, fit=(startlabel)(concern)(define)(construct)(constructlabel), label={[name=evaluate_view_label, font=\small, label distance=2mm]below:\faFileTextO{} \Cref{sec:stimulus}}, label={[font=\itshape\small, label distance=2mm]above:Stimulus}] (section4) {};
  \end{scope}

\end{tikzpicture}
	\caption{Overview of the study: stimulus for evaluation and evaluation with tax experts. Section references indicate where each part is described in the paper.}
	\label{fig:study_overview}
\end{figure*}

This paper contributes 

\begin{enumerate}
	\item a deliberately minimal software architecture viewpoint tailored to tax authorities’ concerns, and
	\item an empirical judgment-based evaluation of the resulting software architecture description and its underlying viewpoint on a large-scale software system conducted with expert tax professionals.
\end{enumerate}

Throughout this paper, we use the terminology and definitions of ISO/IEC/IEEE 42010:2022~\citep{ISO42010}. In particular, we distinguish between architecture description, architecture view, and architecture viewpoint: An \emph{architecture description} is the work product used to express software architectures and is a tangible representation of information provided to stakeholders. An \emph{architecture view} (or simply, \emph{view}) is an information part comprising a portion of an architecture description. It expresses the architecture of the system of interest in accordance with an \emph{architecture viewpoint} (or simply, \emph{viewpoint}), which defines a set of conventions for the creation, interpretation, and use of a view. In this study, the software architecture description under evaluation intentionally comprises a single view constructed using a single, deliberately minimal viewpoint.

The remainder of the article is structured as follows: After establishing the background and related work in \Cref{sec:related_work} and discussing our study design as a judgment study in \Cref{sec:research_method}, we articulate the concerns of tax authorities as stakeholders in globally distributed software architectures (\Cref{sec:concernstakeholder}), define the software architecture viewpoint (\Cref{sec:viewpoint}), and instantiate the view with a large-scale software architecture (\Cref{sec:view}). In \Cref{sec:results}, we present the evaluation results with respect to the minimal viewpoint (\Cref{sec:result_viewpoint}), the resulting view (\Cref{sec:result_view}), and the overall software architecture description (\Cref{sec:result_softwarearchitecturedescription}). After critically reflecting on the evaluation results in \Cref{sec:discussion} and discussing the limitations of our study in \Cref{sec:limitations}, we conclude the paper with a summary of our findings, their implications for research and practice, and directions for future work in \Cref{sec:conclusion}. 

At this point, we emphasize that this work adopts a software-engineering perspective on tax compliance. An in-depth legal or economic analysis is beyond the scope of this study.

\section{Related Work}
\label{sec:related_work}

Since this article connects taxation with software architecture distributed across multiple countries, we provide related work on globally distributed software development and the emerging field of taxation in software engineering, highlighting that existing research does not address software architecture descriptions as artifacts for tax compliance.

\subsection{Globally Distributed Software Development}

Globally distributed software development, particularly in large enterprises, is a common practice driven by cost optimization, resource availability, and global competition \citep{Ruhe2014}. However, cross-border collaboration is considerably more complex than co-located development due to increased communication and coordination overhead. Over time, approaches to managing global projects have evolved from strictly independent work at distributed sites toward more integrated forms of collaboration.

Architectural modularization emerged as an early response to these challenges. Conway’s observation that software architecture reflects organizational structure, known as Conway’s law \citep{Conway1968}, motivated the use of modular architectures to minimize dependencies, allocate work, and reduce coordination needs \citep{Cataldo2008sociotechnical, Herbsleb2003, Herbsleb1999architectures}. While modularization has proven effective, it also has well-documented limitations \citep{Cataldo2008communication, Smite2014, Clerc2007}. Strict separation of work can lead to redundant effort, suboptimal architectural decisions, misplaced functionality \citep{Turecek2010, Smite2014}, and integration problems \citep{Kwan2011, Herbsleb1999splitting}. Moreover, software components are never fully independent, and architectural evolution remains inherently unpredictable \citep{Herbsleb1999splitting}.

Strict code ownership further introduces communication overhead and delays \citep{Clerc2007}. As a result, and supported by agile practices, software development has increasingly shifted toward collaborative approaches. This includes a move from geographically aligned modularization toward componentization using microservices \citep{Newman2021}, as well as the adoption of collective ownership models that allow teams to contribute across component boundaries \citep{Clerc2007}. Collaborative practices such as cross-team coordination, code reviews, and internal reuse of software components \citep{Dorner2022taxing, Smite2023} increase reuse across organizational and national boundaries.

\subsection{Dependency-based Representations in Global Software Development}

Structural dependency representations, such as dependency graphs, coupling metrics, and DSM-like matrices, are widely used in software architecture and global software engineering to analyze coordination needs, modularity, and socio-technical alignment in large-scale systems \citep{Browning2016}. Derived from static or dynamic dependencies between components, these representations support reasoning about architectural decomposition, change impact, and coordination challenges. Our contribution does not lie in proposing a new representation formalism, but in empirically evaluating such representations as external-facing artifacts for a novel stakeholder group, namely tax authorities, for whom these abstractions have not previously been studied.

Existing work primarily targets internal software engineering concerns and assumes architects and developers as the main stakeholders. Dependency-based representations have not yet been empirically studied as external-facing artifacts for non-technical stakeholders \citep{tanveer2023approach}. In particular, their suitability for communicating tax-relevant aspects—such as implicit cross-border licensing, ownership, and jurisdictional distribution—to tax authorities remains unexplored.

In this study, we empirically assess the usefulness of such representations for tax compliance by evaluating them with tax experts.

\subsection{Taxation in Software Engineering}

Taxation in software industry has been debated for many decades \citep{OECD2015}. The problem with taxing the final result of software engineering, the software product or service, for example, has shown to be challenging to tackle and is still subject to ongoing and broad discussion \citep{Olbert2017}. This debate has also reached software engineering, the construction of the software product; at first in the context of inner source, use of open source software development practices and the establishment of open source-like communities within an organization \citep{Capraro2020}. To address the taxation challenges in inner source software development, \citeauthor{Buchner2022} proposed applying a so-called cost-plus approach \citep{Buchner2022}. However, \citeauthor{Treidler2024technicalreport} showed that such a cost-plus approach is not applicable since inner source does not reflect the function and risk profile, the starting point for any transfer pricing discussion, for a cost-plus approach \citep{Treidler2024technicalreport}.

\citeauthor{Dorner2022taxing} identified the use of software components owned by teams or individuals who represent separate geographically distributed subsidiaries of the same enterprise as a type of licensing, a key intangible in software engineering that is taxable~\citep{Dorner2022taxing}. We explain the concerns of tax authorities on software architecture in~\Cref{sec:concernstakeholder}, in laymen's terms and with a simplified example of a multinational enterprise developing software, how tax authorities become a stakeholder in a globally distributed software architecture.

Very recent litigation between software-intensive multinational enterprises and the US tax authorities (IRS), including cases involving Microsoft and Meta (Facebook)\footnote{\url{https://www.taxnotes.com/research/federal/court-documents/court-opinions-and-orders/tax-court-upholds-income-method-facebook-transfer-pricing-case/7s7vd}}, illustrates the importance of tax compliance in software development. For example, in a long-running dispute the IRS alleges that Microsoft owes an additional \$28.9 billion in tax from 2004 to 2013, plus penalties and interest. For the interested reader, we recommend our analysis of the Microsoft case from a legal perspective \citep{Treidler2024tni}.

\section{Research Method}
\label{sec:research_method}

We adopt an evaluation research approach and conduct a \emph{judgment study} to examine to what extent a deliberately minimal software architecture description is perceived by tax experts as making implicit cross-border licensing of multinational enterprises visible and interpretable.

In a judgment study, researchers gather empirical data from a group of participants who are asked to judge or rate behaviors, respond to a request or \emph{stimulus} offered by a researcher, or discuss a given topic of interest \citep{Stol2018}. Stol and Fitzgerald compare judgment studies to a courtroom: a panel of participants (the jury) is carefully and systematically selected, evidence is presented in the form of a stimulus, and the jury eventually returns a verdict \citep{Stol2018}. The setting is only manipulated to the extent that it remains neutral and does not distract participants from the matter at hand. The goal of a judgment study is to achieve generalizability over responses rather than generalizability to a population of actors \citep{Stol2018}. Consequently, a small, carefully selected panel of domain experts is appropriate for this type of evaluation.

In our study, a software architecture description of a large industrial software system serves as the evaluation stimulus. The stimulus comprises an architecture viewpoint and a single architecture view constructed in accordance with that viewpoint, together forming the software architecture description under evaluation. The experts’ assessments of this stimulus, collected through semi-structured interviews, constitute the empirical data of the study.

This methodological choice is motivated by the interpretive nature of tax compliance and transfer pricing, which rely heavily on professional expertise and contextual reasoning. Since tax authorities are not a traditional stakeholder group in software architecture, expert judgment is required to assess whether architectural abstractions are meaningful and usable in this context. Accordingly, the goal of this study is analytical generalization, i.e., understanding how tax experts reason about architectural information, rather than statistical generalization to a broader population. We do not aim to validate a complete tax compliance solution or propose a prescriptive architectural method; instead, we evaluate how experts reason about architectural information when presented with a deliberately scoped evaluation stimulus.

\subsection{Expert Panel and Sampling}

The judgment study was conducted with a panel of four senior tax experts representing the two perspectives typically involved in a tax audit: the tax authority perspective and the company-side advisory perspective. The panel therefore comprised two tax auditors and two tax advisors with expertise in transfer pricing, thereby covering both the recipients and the providers of tax-relevant documentation.

Participants were selected through purposeful sampling from our professional network based on three criteria: 

\begin{itemize}
	\item Extensive professional experience in international taxation and transfer pricing,
	\item direct involvement in tax audits of multinational enterprises, and
	\item familiarity with the taxation of intangible assets. 
\end{itemize}

All participants had multiple years of professional experience in their respective roles. \Cref{tab:interviewees} provides an overview of our interviewees and their backgrounds.

\begin{table}[H]
  \centering
  \small
  \renewcommand{\arraystretch}{1.2}
  \caption{An overview of the interviewees and their backgrounds.}
  \label{tab:interviewees}
  \begin{adjustwidth}{-\extralength}{0cm}
  \begin{tabularx}{\linewidth}{L{1em}L{2cm}X}
    \toprule
    {ID} & {Role} & {Background} \\ \midrule
    A & Tax auditor & Former tax auditor, majoring in international taxation, OECD advisor for transfer pricing from 2016 to 2018, currently not involved in transfer pricing \\
    K & Tax advisor & Transfer pricing advisor, previously transfer pricing expert at one of the Big Four accounting firms for 20~years \\
    M & Tax auditor & Tax auditor and key expert on transfer pricing for other tax auditors for more than eight years \\
    R & Tax advisor & Managing director at a transfer price consulting firm; tax inspector before, 35~years of experience in international taxation \\  \bottomrule
  \end{tabularx}
  \end{adjustwidth}
\end{table}

None of the participants were compensated for their participation. None had prior academic or professional collaboration with members of the author team. The composition of the panel is consistent with established practice in judgment studies, where depth of expertise and diversity of perspectives are prioritized over statistical representativeness.

\subsection{Evaluation Stimulus}

The evaluation stimulus is a software architecture description constructed to elicit expert judgment on tax-relevant aspects of globally distributed software architectures. The software architecture description consists of two elements: an architecture viewpoint that addresses the concerns of tax authorities and an instantiated architecture view derived from a large-scale industrial software system.

The architecture view represents the software system as a set of software components and their dependencies, enriched with information about component ownership and the jurisdictions of the owning organizational units. Dependencies capture structural usage relationships between components across organizational and geographical boundaries. The stimulus focuses on architectural information relevant to identifying potential instances of implicit cross-border licensing.

The view was instantiated using historical architectural and organizational data extracted from an internal company system that aggregates information from code repositories and team registries. Only software components deployed in production were included, as these directly or indirectly contribute to customer-facing value creation. Components owned by individuals rather than organizational units were excluded due to privacy constraints.

To preserve confidentiality, all company-identifying information was removed, geographic distributions were altered, and the case company is not identifiable. The stimulus is intentionally scoped and incomplete by design: it does not aim to represent the full software architecture, capture runtime behavior, quantify economic value, or support operational tax compliance. Its sole purpose is to support expert evaluation of the suitability and limitations of architectural descriptions for tax-related reasoning.

We operationalize internal reuse as cross-component dependency relations, as these constitute a traceable technical proxy for cross-entity use.

\subsection{Data Collection}

We collected empirical data through semi-structured interviews with each expert individually. Prior to the main study, we conducted two pilot interviews within the author team and one additional pilot interview with a tax auditor who was not part of the expert panel. These pilot interviews were used to address didactical challenges arising from the experts’ taxation background and to refine the interview flow and explanations of software engineering concepts.

The interviews were conducted remotely between October~24 and November~3,~2023. Each interview was facilitated by a pair of researchers: one with a software engineering background and one with a taxation background. This pairing ensured accurate explanation of architectural concepts and appropriate contextualization of tax-related discussions. All interviews were recorded and conducted in English or German.

The interviews focused on the experts’ assessment of the evaluation stimulus, including the perceived plausibility and adequacy of the architecture viewpoint, the usefulness of the architecture view for tax audits, the interpretability of the architectural abstractions, and perceived limitations and risks. The interview guidelines are publicly available \citep{Dorner2026}.

\subsection{Data Analysis}

We analyzed the interview data using a reflexive form of thematic analysis \citep{Clarke2017}. The analysis focused on the experts’ judgments regarding the plausibility, interpretability, usefulness, and limitations of the evaluation stimulus. Given that substantial parts of the interviews were dedicated to explaining software engineering concepts, we translated and paraphrased only the experts’ substantive responses relevant to these questions.

The first author conducted the initial coding inductively, first familiarizing himself with the interview recordings and transcripts and then assigning descriptive codes to statements concerning the architecture viewpoint, the instantiated view, and the overall software architecture description. Codes were iteratively refined as additional material was analyzed and were subsequently grouped into broader themes based on conceptual and semantic similarities.

The second author reviewed the resulting codes, themes, and their mappings in a separate validation session. Consistent with reflexive thematic analysis, we did not seek inter-rater reliability or independent parallel coding, but used collaborative discussion to challenge interpretations and refine the thematic structure. 

To mitigate translation errors and misunderstandings, the final version of the paper was shared with all interview participants for confirmation of factual accuracy and intent.

\subsection{Ethical and Confidentiality Considerations}

Given the sensitivity of both tax-related and architectural information, we ensured strict privacy and confidentiality throughout the study. All interviews were conducted individually, and all company-specific information was removed from the data and the presented material. To further prevent identification of the case company, we altered the geographical distribution of subsidiaries while preserving the characteristics of a globally distributed software architecture.

Participants were informed about the purpose of the study and the use of the collected data. None of the participants were compensated. Apart from the publicly available interview guidelines, no raw architectural data or interview transcripts are shared.

\section{A Software Architecture Description for Tax Compliance}
\label{sec:stimulus}

This section introduces the software architecture description used as the evaluation stimulus in our judgment study. It comprises a single stakeholder concern, a corresponding architecture viewpoint, and a single architecture view constructed in accordance with that viewpoint for a large-scale industrial software system. The purpose is not to determine transfer prices, but to assess whether the software architecture description makes tax-relevant cross-border software reuse visible and interpretable to tax experts.

\subsection{Concern: International Taxation and Software Architecture}
\label{sec:concernstakeholder}

From an international taxation perspective, multinational enterprises are required to identify and document cross-border transactions involving intangibles, including software. While software code itself is intangible and not bound to a physical location, its ownership and control are legally attributed to specific entities located in specific jurisdictions. When software components developed or owned by one legal entity are used by another entity within the same corporate group, this use constitutes an often implicit form of licensing that may be taxable under international transfer pricing regulations.

To illustrate this concern, consider the fictional \emph{devnullsoft Group}, a multinational enterprise that develops and sells a software product. The product is decomposed into multiple software components, each owned by a team that is responsible and accountable for its development, maintenance, and operation. The teams are assigned to different legal entities of the devnullsoft Group based on their location: devnullsoft GmbH in Germany, devnullsoft AB in Sweden, and devnullsoft Ltd.\ in the UK. \Cref{fig:overview} provides a schematic overview of this simplified globally distributed software architecture, showing components, their owning entities, and dependencies between components across jurisdictions.

\begin{figure*}
  \centering
  \begin{tikzpicture}[
  x=1.2cm, y=1.2cm,
      component/.style={rectangle, draw, inner sep=2mm, outer sep=0pt, fill=white}
    ]

	\node[component] at (0,0) (c00) {};
	\node[component] at (0,1) (c01) {};
	\node[component] at (0,2) (c02) {};
	
	\node[component] at (1,0) (c10) {};
	\node[component] at (1,1) (c11) {};
	\node[component] at (1,2) (c12) {};
	
	\node[component] at (4,0) (c30) {};
	\node[component] at (4,1) (c31) {};
	\node[component] at (4,2) (c32) {};
	
	\node[component] at (5,0) (c40) {};
	\node[component] at (5,1) (c41) {};
	\node[component] at (5,2) (c42) {};
	
	\node[component] at (6,0) (c50) {};
	\node[component] at (6,1) (c51) {};
	\node[component] at (6,2) (c52) {};
	
	\node[component] at (10,0) (c70) {};
	\node[component] at (10,1) (c71) {};
	\node[component] at (10,2) (c72) {};

    \foreach \fromx/\fromy/\tox/\toy/\b/\c in {%
      4/1/7/1/left/ACMRed,%
      3/2/1/1/right/ACMRed,%
      1/1/3/0/left/ACMRed,%
      3/1/1/2/right/ACMRed,%
      3/1/4/0/left/gray,%
      0/2/0/1/right/gray,%
      1/0/0/1/right/gray,%
      1/0/3/0/left/ACMRed,%
      1/1/0/1/right/gray,%
      1/2/0/2/right/gray,%
      5/2/7/2/left/ACMRed,%
      5/2/7/1/left/ACMRed,%
      5/1/7/1/left/ACMRed,%
      5/0/7/1/left/ACMRed,%
      7/1/7/0/left/gray,%
      4/2/3/2/right/gray,%
      4/2/5/2/left/gray%
    } \draw[-stealth, thick, color=\c] (c\fromx\fromy) to[bend \b] (c\tox\toy);


    \begin{scope}[on background layer, y=1cm]

      \node[draw, rounded corners=1em, fit={(c00) (c12)}, rectangle, inner sep=1em, outer sep=0em, fill=black!5] (ownerab) {};

      \node[draw, rounded corners=1em, fit={(c30) (c52)}, rectangle, inner sep=1em, outer sep=0em, fill=black!5] (ownergmbh) {};

      \node[draw, rounded corners=1em, fit={(c70) (c72)}, rectangle, inner sep=1em, outer sep=0em, fill=black!5] (ownerltd) {};

      \node[anchor=south, draw, above=1.2cm of ownerab.north, minimum height=2em] (ab) {devnullsoft AB};
      \node[anchor=south, above=1.2cm of ab.north] (swedishtax) {\worldflag[width=0.5cm, framewidth=0mm]{SE}};
      \draw[-stealth] (ab) -- (swedishtax) node[midway, fill=white] {is taxable in};
      \draw[-stealth] (ab) -- (ownerab.north) node[midway, fill=white] {owns};
      
      \node[anchor=south, draw, above=1.2cm of ownergmbh.north, minimum height=2em] (gmbh) {devnullsoft GmbH};
      \node[anchor=south, above=1.2cm of gmbh.north] (germantax) {\worldflag[width=0.5cm, framewidth=0mm]{DE}};
      \draw[-stealth] (gmbh) -- (germantax) node[midway, fill=white] {is taxable in};
      \draw[-stealth] (gmbh) -- (ownergmbh.north) node[midway, fill=white] {owns};

      \node[anchor=south, draw, above=1.2cm of ownerltd.north, minimum height=2em] (ltd) {devnullsoft Ltd.};
      \node[anchor=south, above=1.2cm of ltd.north] (uktax) {\worldflag[width=0.5cm, framewidth=0mm]{GB}};
      \draw[-stealth, thick, ] (ltd) -- (uktax) node[midway, fill=white] {is taxable in};
      \draw[-stealth, thick, ] (ltd) -- (ownerltd.north) node[midway, fill=white] {owns};

      \draw[stealth-stealth, thick, ACMRed, dashed] (ab) -- (gmbh) node[midway, text centered, above] {Transfer price};
      \draw[stealth-stealth, thick, ACMRed, dashed] (gmbh) -- (ltd) node[midway, text centered, above] {Transfer price};

      


    \end{scope}


  \end{tikzpicture}
  \caption{A schematic overview of a globally distributed software architecture developed by the fictional \emph{devnullsoft Group}. Software components \softwarecomponenttikz{} are owned by different legal entities, and cross-jurisdictional dependencies \dependencytikz{} illustrate internal use relationships that are relevant for transfer pricing.}
  \label{fig:overview}
\end{figure*}

Only devnullsoft GmbH in Germany sells the final software product to customers. Without further consideration, all profits would therefore be taxed exclusively in Germany. The Swedish and British subsidiaries, however, develop, maintain, and own software components that are reused by the product sold by the German entity. From the perspective of the Swedish and UK tax authorities, this situation is problematic, as value-contributing activities take place in their jurisdictions without an associated share of taxable profit.

To avoid this scenario and to provide a common ground for international taxation, reducing uncertainty for multinational enterprises, and preventing tax avoidance through profit shifting, nearly all countries in the world agreed on and implemented the so-called \emph{arm's length principle} as defined in the \emph{OECD Transfer Pricing Guidelines for Multinational Enterprises and Tax Administrations}~\citep{OECD2022}. This has become the guiding principle and the de-facto standard for the taxation of multinational enterprises that requires associated enterprises to operate as if not associated and regular participants in the market from a taxation perspective. To comply with the arm's length principle, multinational enterprises must pay a so-called \emph{transfer price}. Transfer prices are the prices at which an enterprise transfers physical goods or---for software much more relevant---intangibles or provides services to associated enterprises. Those transfer prices are established on a market value basis and aim to avoid profit shifts from high to low tax regions. Although other types of intangibles, like cross-border code contributions, code reviews, bug reports, etc., are also relevant from a taxation point of view \citep{Dorner2022taxing}, we focus in this article on another highly relevant type of intangible in software engineering: licenses.

Although software licenses are often associated with external or open-source software, the internal use and reuse of software components within a multinational enterprise also constitutes licensing from a taxation perspective. Even when such licenses are not explicitly defined, the exclusion of third parties from using internally developed software components implies that these components are controlled and licensed within the corporate group. Without appropriate documentation and transfer pricing arrangements, this situation contradicts the arm’s length principle and renders multinational enterprises non-compliant with international taxation standards.

In this way, tax authorities become stakeholders in globally distributed software architectures as they have the concern of identifying cross-border reuse of software components as an instance of taxable licensing. Determining the actual transfer pricing for the cross-border reuse of software components is a vast and separate challenge on its own and not part of this work. Our work focuses on the architectural description comprising all necessary information for reporting the software architecture to tax authorities.

From this concern, we derive three fundamental questions that a software architecture description must be able to answer to support tax compliance: 

\begin{enumerate}
  \item How is the software-intensive product structured?
	\item Which organizational owners are associated with these components and their cross-border reuse relationships?
  \item Where are those legal entities geographically located?
\end{enumerate}

All three questions must be answered by a software architecture description to ensure tax compliance for multinational enterprises.

\subsection{Architecture Viewpoint for Tax-Related Concerns}
\label{sec:viewpoint}

To address the concern of tax authorities and to answer the three derived questions, we define a software architecture viewpoint that specifies the conventions for creating, interpreting, and using an architecture view for tax-related concerns.

The viewpoint requires four elements: First, the notion of a software component must be defined. A software component is understood as a self-contained, reusable unit that encapsulates functionality and can be developed, maintained, and used independently. The concrete definition of a component is company-specific and must be justified in the context of reporting.

Second, component ownership must be defined. Each software component must be assigned to exactly one organizational owner that is responsible and accountable for the component. Although different notions of code ownership exist in software engineering \citep{Fowler2006, bird2011don, greiler2015code, Thongtanunam2016, posnett2013dual, Zabardast2022}, this viewpoint requires a single accountable owner for reporting purposes. The chosen definition of ownership must be documented and justified.

Third, the component structure of the software product must be decomposed to identify dependencies between components. These dependencies reveal which components use other components and form the basis for identifying potential instances of implicit licensing.

Fourth, the jurisdiction of each component owner must be identified. In most cases, the jurisdiction corresponds to the country in which the owning legal entity is located.

Together, these elements define a viewpoint that enables the construction of an architecture view exposing the implicit cross-border licensing of a globally distributed software system.

\begin{table}[H]
\centering
\small
\renewcommand{\arraystretch}{1.2}
\caption{Operational specification of the minimal architecture viewpoint.}
\label{tab:viewpoint_spec}
\begin{adjustwidth}{-\extralength}{0cm}
\begin{tabularx}{\linewidth}{L{3cm} X X}
\toprule
\textbf{Element} & \textbf{Definition} & \textbf{Instantiation in this study} \\
\midrule
Software component & A self-contained, reusable unit of functionality & Microservices deployed in production \\
Owner & Exactly one organizational unit responsible and accountable for the component & Team registered as responsible for the microservice \\
Jurisdiction & Country associated with the owner for tax purposes & Country self-reported in the internal team registry; may be unknown or ambiguous \\
Dependency & Directed usage relation between two components & Extracted from the internal architectural data source \\
Aggregation & Rule for summarizing dependencies & Dependencies aggregated from component user jurisdiction to component owner jurisdiction \\
Exclusions & Elements intentionally left out of scope & Non-production services, individually owned services, and runtime/economic information \\
\bottomrule
\end{tabularx}
\end{adjustwidth}

\end{table}

\subsection{Architecture View: Industrial Case Application}
\label{sec:view}

We construct a single architecture view in accordance with the viewpoint for a large-scale industrial software system developed by a multinational enterprise. This view represents a snapshot of the architecture at a specific point in time. 

Due to confidentiality constraints, we cannot describe the case company and data collection in full detail to maintain the anonymity of our case company, which was used as a source for creating our view. However, the examined system exhibits characteristics commonly found in multinational enterprises developing large-scale software systems. 

In this case, microservices are used as the unit of decomposition, as they represent independently deployable and self-contained units of functionality \citep{Newman2021} with clear ownership boundaries. The software product consists of 2{,}560 microservices in production. Experimental and non-production services are excluded, as only production components contribute directly or indirectly to customer-facing functionality.

Each microservice is owned by a team that is responsible and accountable for its development, operation, and maintenance. We deliberately use an operational notion of ownership for architectural reporting, acknowledging that this differs from legal ownership used in taxation. In addition, 42 microservices (1.64\%) owned by individual developers were excluded from the analysis, as the geographical location of individuals could not be accessed due to privacy constraints. This exclusion does not affect the conceptual validity of the viewpoint but reflects practical limitations of data availability in industrial settings. After filtering, the resulting architecture comprises 2{,}518 microservices owned by 336 teams and connected by 16{,}533 dependency relationships.

Teams self-report their location in an internal organizational system. While this reflects common practice in distributed software development, it results in incomplete jurisdictional information for some teams, particularly those that are fully remote or geographically distributed. As a consequence, not all component owners can be unambiguously assigned to a single jurisdiction, leading to dependencies with unknown or ambiguous jurisdictional assignments.

The resulting architecture view is presented using two complementary representations. \Cref{fig:graph} shows a directed graph-based visualization in which nodes represent jurisdictions and directed edges represent cross-jurisdictional use relationships between software components. An edge is directed from the component user to the component owner, reflecting the direction of implicit licensing from a taxation perspective. Edge weights correspond to the number of such dependency relationships aggregated between jurisdictions. Dependencies involving owners with unknown or ambiguous jurisdictions are omitted from the graph to preserve readability. A corresponding tabular representation complements the graph by explicitly capturing all dependencies, including those involving unknown or ambiguous jurisdictional assignments. Together, these representations form the software architecture description evaluated by the panel of tax experts in \Cref{sec:results}.

\begin{figure}
  \centering
  \begin{tikzpicture}[%
      country/.style={circle, draw, minimum width=2em},
      use/.style={-latex, >=latex}%
    ]

    \node[country, label={right:Germany}] at (1*360/5: 3.5cm)  (DEU) {};
    \node[country, label={left:USA}] at (2*360/5: 3.25cm)  (USA) {};
    \node[country, label={left:France}] at (3*360/5: 3.25cm)  (SWE) {};
    \node[country, label={right:Netherlands}] at (4*360/5: 3.25cm)  (NLD) {};
    \node[country, label={below right:UK}] at (5*360/5: 3.25cm)  (GBR) {};

    \draw[use] (USA) to[bend left=12] node[midway, fill=white] {1130} (SWE);
    \draw[use] (GBR) to[bend left=10] node[midway, fill=white] {261} (SWE);
    \draw[use] (SWE) edge[out=310, in=220, loop] node[midway, fill=white] {4069} (SWE);
    \draw[use] (USA) edge[loop] node[midway, fill=white] {1648} (USA);
    \draw[use] (SWE) to[bend left=12] node[midway, fill=white] {850} (USA);
    \draw[use] (GBR) to[bend left=10] node[midway, fill=white] {43} (USA);
    \draw[use] (NLD) to[bend left=10] node[midway, fill=white] {11} (SWE);
    \draw[use] (GBR) edge[out=90, in=0, loop] node[midway, fill=white] {164} (GBR);
    \draw[use] (SWE) to[bend left=10] node[midway, fill=white] {108} (GBR);
    \draw[use] (USA) to[bend left=10] node[midway, fill=white] {24} (GBR);
    \draw[use] (SWE) to[bend left=10] node[midway, fill=white] {24} (DEU);
    \draw[use] (USA) to[bend left=10] node[midway, fill=white] {14} (DEU);
    \draw[use] (GBR) to[bend left=10] node[midway, fill=white] {15} (DEU);
    \draw[use] (NLD) to[bend left=10] node[midway, fill=white] {3} (DEU);
    \draw[use] (DEU) edge[loop] node[midway, fill=white] {2} (DEU);
    \draw[use] (NLD) to[bend left=10] node[midway, fill=white] {6} (GBR);
    \draw[use] (DEU) to[bend left=10] node[midway, fill=white] {2} (GBR);
    \draw[use] (NLD) to[bend left=10] node[midway, fill=white] {5} (USA);
    \draw[use] (SWE) to[bend left=10] node[midway, fill=white] {21} (NLD);
    \draw[use] (USA) to[bend left=10] node[midway, fill=white] {15} (NLD);
    \draw[use] (NLD) edge[out=310, in=220, loop] node[midway, fill=white] {19} (NLD);
    \draw[use] (GBR) to[bend left=10] node[midway, fill=white] {2} (NLD);

  \end{tikzpicture}
  \caption{Geographical distribution of the microservice architecture of the case company using the proposed architectural viewpoint. Arrows indicate dependencies from microservice users to microservice owners; country labels are partially substituted to preserve confidentiality.}
  \label{fig:graph}
\end{figure}
This graph, however, lacks information about unclear or ambiguous locations. Therefore, we also provide a tabular form of the graph, including the information on missing or unclear locations. \Cref{tab:table} depicts this information.
\begin{table}
  \centering
  \renewcommand{\arraystretch}{1.2}
    \caption{An equivalent tabular representation of \Cref{fig:graph} but including unknown dependencies with unknown or ambiguous jurisdictional assignments.}
  \begin{tabular}{cc|cccccc}
    & & \multicolumn{6}{c}{\textbf{Component owner}}\\
    & & DEU & GBR & NLD & FRA & USA & N/A \\ \hline
    \multirow{6}{*}{\rotatebox[origin=c]{90}{\textbf{Component user}}} & DEU & 2 & 2 & 0 & 0 & 0 & 4 \\
    & GBR & 15 & 164 & 2 & 261 & 43 & 141 \\
    & NLD & 3 & 6 & 19 & 11 & 5 & 8 \\
    & FRA & 24 & 108 & 21 & 4069 & 850 & 1767 \\
    & USA & 14 & 24 & 15 & 1130 & 1648 & 642 \\
    & N/A & 27 & 70 & 14 & 2283 & 970 & 2171 \\
  \end{tabular}
  \label{tab:table}
\end{table}

\section{Evaluation Results}
\label{sec:results}

This section reports the experts' judgments on the evaluation stimulus with respect to the visibility, interpretability, and perceived usefulness of the software architecture description, as well as its perceived limitations. \Cref{tab:themes} summarizes the main themes identified across the interviews and indicates which interviewees explicitly raised them. These coverage indications are provided to support traceability of the thematic analysis and to distinguish broad agreement from differences in emphasis; they are not intended as quantitative evidence.

\newcommand{\yesbox}{\(\blacksquare\)}
\newcommand{\nobox}{\(\square\)}
\newcolumntype{Y}{>{\centering\arraybackslash}p{1.6em}}

\begin{table}[H]
  \centering
  \caption{Overview of the main themes identified in the expert judgments and their coverage across interviewees. Filled boxes indicate that a theme was explicitly raised by the respective interviewee; empty boxes indicate that it was not explicitly raised. The table is intended to support traceability of the thematic analysis and to distinguish broad agreement from differences in emphasis, rather than to provide quantitative evidence. Interviewees A and M are tax auditors; K and R are tax advisors.}
  \label{tab:themes}
  \begin{adjustwidth}{-\extralength}{0cm}
  \renewcommand{\arraystretch}{1.2}
  \small
  \begin{tabularx}{\fulllength}{p{3.5cm} Y Y Y Y X}
    \toprule
    \textbf{Theme} & \textbf{A} & \textbf{K} & \textbf{M} & \textbf{R} & \textbf{Summary} \\
    \midrule
    Viewpoint is a plausible starting point 
      & \yesbox & \nobox & \nobox & \yesbox 
      & The minimal viewpoint was judged as a useful starting point for making implicit licensing structures visible. \\
    
    Viewpoint insufficient for valuation 
      & \nobox & \yesbox & \nobox & \yesbox 
      & The viewpoint was consistently considered insufficient for valuation because components are not additive and dependency counts do not express economic relevance. \\
    
    Ownership concept mismatch 
      & \nobox & \yesbox & \yesbox & \nobox 
      & Experts interpreted ownership primarily in legal terms, revealing a mismatch with the software-engineering notion of ownership used in the viewpoint. \\
    
    Jurisdictional aggregation is context-dependent 
      & \nobox & \yesbox & \nobox & \nobox 
      & Aggregation by jurisdiction was considered acceptable in the presented case, but not generally applicable in more complex multinational structures. \\
    
    Jurisdictional ambiguity weakens auditability 
      & \yesbox & \nobox & \yesbox & \nobox 
      & Missing or ambiguous owner--jurisdiction assignments were seen as one of the main limitations of the constructed view. \\
    
    Traceability of assignments is important 
      & \yesbox & \nobox & \yesbox & \nobox 
      & Especially the tax auditors emphasized that audits require traceable assignment procedures, not only final labels. \\
    
    Graph is useful for high-level orientation 
      & \yesbox & \nobox & \yesbox & \nobox 
      & The graph representation was positively received as a useful entry point for discussion and exploration. \\
    
    Table complements the graph 
      & \nobox & \nobox & \yesbox & \nobox 
      & The tabular representation was seen as a useful complement because it preserves absolute counts and ambiguous assignments. \\
    
    Dependency counts are not value proxies 
      & \nobox & \yesbox & \nobox & \yesbox 
      & Experts emphasized that weighted dependencies should not be interpreted as indicators of value or relevance. \\
    
    Qualitative distinctions between components are missing 
      & \nobox & \yesbox & \nobox & \yesbox 
      & The instantiated view lacks qualitative information needed to distinguish replaceability, differentiation, or business significance. \\
    
    Snapshot and exclusions limit representativeness 
      & \nobox & \nobox & \yesbox & \yesbox 
      & Experts questioned the exclusion of non-production components and the representativeness of a single architectural snapshot. \\
    
    Software architecture description provides an objective starting point 
      & \yesbox & \yesbox & \yesbox & \yesbox 
      & The overall software architecture description was judged as a credible, evidence-based starting point for tax audit discussions. \\
    
    Useful for discussion, not sufficient for tax assessment 
      & \nobox & \yesbox & \yesbox & \yesbox 
      & The software architecture description was considered useful for structuring discussions, but insufficient on its own for legal interpretation or transfer pricing decisions. \\
    \bottomrule
  \end{tabularx}
  \end{adjustwidth}
\end{table}

\subsection{Architecture Viewpoint}
\label{sec:result_viewpoint}

The experts consistently judged the minimal viewpoint as a plausible and relevant starting point for framing the tax-related concern. Across the interviews, the panel agreed that the three questions derived from the concern of international taxation are meaningful and that the viewpoint makes a substantial part of the implicit licensing structure visible. As interviewee A stated: \myenquote{I think it is possible to reveal the licensing structure [by your viewpoint], at least by 90\%, there will always be fuzziness.}

At the same time, the experts consistently emphasized that the viewpoint is not sufficient as a standalone basis for tax assessment. A first recurring theme concerned the distinction between identifying potentially relevant software relationships and supporting valuation. The experts agreed that the viewpoint helps identify who is involved, but does not provide the information needed to reason about the economic relevance of components. As interviewee R said: \myenquote{You are approaching it from a taxation perspective from a very detailed level}. The same interviewee continued: \myenquote{[Your viewpoint] is an important first step towards understanding who is actually contributing to the software product. But to answer the follow-up question, which is the valuation of the software component, you need to add another layer. [Components] are not additive.} Interviewee K expressed the same concern succinctly: \myenquote{Value is not counting components.} Taken together, these statements indicate consensus that the viewpoint is useful for surfacing structural relations, but insufficient for reasoning about value without additional qualitative distinctions.

A second recurring theme concerned ownership. Multiple experts, particularly interviewees M and K, interpreted ownership primarily in a legal sense and linked it to accountability from a tax perspective. They raised the question of who is actually accountable for the software product or the components from a legal perspective. This revealed a systematic mismatch between the software-engineering notion of ownership used in the viewpoint and the legal or economic notions of ownership relevant in taxation. The issue was therefore not a rejection of the viewpoint's structure, but a recurring reinterpretation of one of its core concepts through a legal lens.

A third theme concerned jurisdictional representation. Several experts accepted aggregation by jurisdiction in the presented case, but considered it a context-specific simplification rather than a general solution. The panel emphasized that aggregation at country level is only meaningful as long as one jurisdiction corresponds to one relevant subsidiary. Interviewee K highlighted that assigning legal entities to jurisdiction may become complicated in cases involving so-called \emph{permanent establishments}: \myenquote{Assigning jurisdiction is complex. There are gray areas.} This indicates that the viewpoint's geographical abstraction was judged as acceptable for an initial structural representation, but too coarse for all realistic taxation settings.

Overall, the panel judged the viewpoint as a plausible and relevant starting point for framing tax concerns. At the same time, the experts consistently identified structural limitations concerning legal ownership, jurisdictional interpretation, and valuation, indicating that the viewpoint is useful for orientation but not sufficient as a standalone basis for tax assessment.

\subsection{Architecture View}
\label{sec:result_view}

The experts considered the instantiated view informative and useful, but they also identified recurring limitations that affected its interpretability and auditability. Across the interviews, two dominant themes emerged: unclear jurisdictional assignments and the limited meaning of purely quantitative dependency counts. In addition, the auditors and advisors emphasized different practical concerns: the tax auditors focused primarily on traceability and completeness, whereas the advisors focused more strongly on valuation and explanatory richness.

The most frequently emphasized issue concerned unclear or ambiguous links between owners and jurisdictions. The view contained a substantial share of dependencies for which jurisdictional assignment was missing or ambiguous, and the experts consistently considered this problematic. Interviewee M positively assessed the underlying company-internal team platform as a data source: \myenquote{I go along with that one. [...] That is, in my opinion, an excellent data source}. At the same time, both tax auditors stressed that the remaining ambiguity was too extensive for audit purposes. Interviewee M stated that \myenquote{missing some percentages is no problem}, but that \myenquote{half of the dependencies missing is too much}. Interviewee A expressed the same concern more strongly: \myenquote{50\% not assignable implies that 50\% of all transfer prices are not properly assigned. [...] Without data, we are in a vacuum.} These judgments show a clear consensus that the quality of the data source is not the core issue; rather, the unresolved jurisdictional assignments substantially weaken the view's usefulness for tax reasoning.

Related to this, the tax auditors emphasized traceability. Interviewees M and A stressed that a tax audit would require not only the assignments themselves, but also visibility into the process by which those assignments were established. They suggested additions such as the jurisdiction of the team manager, cascades over additional organizational data sources, or questionnaires to component owners to improve traceability and reduce ambiguity. In this context, the panel agreed that the physical location of developers is of secondary importance compared to the tax-relevant organizational assignment of ownership and accountability.

A second strong theme concerned the presented graph and table as useful but limited representations. The graph visualization was positively received, especially as a high-level orientation mechanism. Interviewee M said: \myenquote{I like the [graph] visualization; it is helpful}. Interviewee A described it as \myenquote{helpful for a high-level view, a good starting point for looking into details}. The tabular representation was then seen as a useful complement because it preserved dependencies with unknown or ambiguous jurisdictional assignments. Interviewee M explicitly preferred absolute over relative numbers, noting that \myenquote{relative numbers are easy to calculate} anyway. Together, these statements indicate that the experts found the view operationally useful as a structured entry point into discussion, especially when the graph and table were interpreted together.

At the same time, the panel consistently rejected any interpretation of the quantitative dependency counts as indicators of value. This concern was expressed most strongly by the advisors, but was consistent with the broader panel discussion. The experts emphasized that the weighted edges should be understood only as counts of technical dependencies, not as evidence of economic relevance or value creation. As already indicated in the discussion of the viewpoint, the assumption that all components are equal was seen as untenable. Interviewee R summarized this directly: \myenquote{The edge weight does not reflect the relevance}. Interviewee K added that \myenquote{Value is not counting components.} These judgments indicate that the view succeeds in visualizing structural connectedness, but not in expressing tax-relevant distinctions between different kinds of components.

A further recurring theme was the absence of qualitative and contextual information. According to the experts, the quantitative view lacks the distinctions needed to judge replaceability, differentiation, or business significance. This concern was closely related to the more general issue that the architecture view represents structure, but not the characteristics that would be needed for downstream valuation discussions.

Finally, the panel raised three additional limitations of the instantiated view. First, interviewee R questioned whether the effort required to generate the view would be justified in practice and emphasized that the cost-benefit ratio must remain positive. Second, interviewee M criticized the exclusion of non-production and experimental components, noting that such components may still have significant value and that excluding them reduces explanatory power. Third, interviewee M questioned the representativeness of a single architectural snapshot and raised the issue of temporal stability from the perspective of tax audits covering fiscal periods rather than single points in time.

Overall, the experts judged the instantiated view as informative and practically useful for making cross-border software reuse discussable, especially through the combined use of graph and table. However, they also consistently identified limitations related to ambiguous jurisdictional assignments, the lack of qualitative distinctions, and the risk of misinterpreting dependency counts as indicators of value.

\subsection{Software Architecture Description}
\label{sec:result_softwarearchitecturedescription}

When asked for their overall assessment of the software architecture description in its current form, the experts expressed a broadly positive view of its role as a structured, evidence-based basis for discussion. Across the interviews, a clear consensus emerged that the description provides a more objective starting point for tax audit discussions than ad hoc or purely narrative accounts of software development structures.

This positive overall judgment was expressed most explicitly by interviewee R, who summarized the role of the software architecture description as follows: \myenquote{This is just fact-driven, it is no interpretation. When you ask people, they may or may not know who is involved. The nice thing about it is that it is not a subjective view; it is about facts, it is about technical details. This is what is used, if you like it or not. That is relevant because you have objectivized the contributions [to the software product]. Those are the parties involved. This allows you to have a very objective starting point to consider the potential parties you have to consider in greater detail.} This judgment was consistent with the way the other experts described the value of the software architecture description: not as a complete compliance solution, but as a credible and structured basis for more informed discussion.

A second recurring theme concerned the productive role of uncertainty. Rather than criticizing the software architecture description for not eliminating all ambiguity, the experts acknowledged that uncertainty is inherent to tax audits and appreciated that the software architecture description makes some of that uncertainty visible. Interviewee R summarized this broader perspective by stating that \myenquote{there is no black or white in tax audits but a lot of gray}. In this sense, the software architecture description was seen as useful not because it resolves all interpretive problems, but because it surfaces relevant parties, dependencies, and unresolved issues in a more explicit way.

The panel also consistently affirmed the relevance of software architecture as the right level of abstraction for understanding how software products are composed and reused across organizational boundaries. Interviewee K stated: \myenquote{I think software architecture is the right perspective to understand how software is composed conceptually}. Interviewee R additionally highlighted the broader practical need for software support to manage this complexity in transfer pricing processes. Interviewee M, from the tax authority perspective, emphasized the unusual quality of the presented documentation in comparison to ordinary audit practice: \myenquote{A company with such detailed and thorough documentation would be among the top 10 or top 5 percent of what a tax auditor usually sees.} Interviewee K expressed a similar appreciation for the attempt to objectify the discussion: \myenquote{I really like the approach aiming for objectiveness. Let's face it: Sometimes, there are preposterous debates in tax audits.}

At the same time, this positive overall assessment did not eliminate the limitations identified at the viewpoint and view levels. The experts did not regard the software architecture description as sufficient for determining transfer prices or for deriving legally meaningful conclusions about ownership, jurisdiction, or value. Instead, they consistently positioned it as a useful starting point that would need to be complemented by additional legal, organizational, and economic information.

Overall, the panel judged the software architecture description as a credible, objective, and practically valuable basis for initiating tax-related analysis of globally distributed software systems. Its primary perceived usefulness lies in making potentially relevant cross-border reuse relationships visible and discussable, while its main limitation lies in the inability of software architecture abstractions alone to support the legal and economic interpretations required for tax assessment.

\section{Discussion}
\label{sec:discussion}

In this section, we interpret the evaluation results in light of our research objective of assessing to what extent a deliberately minimal software architecture description can make implicit cross-border licensing visible and interpretable for tax experts, and where software architecture abstractions remain limited for tax-related reasoning.

A central finding of our evaluation is the high proportion of ambiguous assignments between component owners and jurisdictions, affecting approximately half of all cross-component dependencies. While this degree of ambiguity poses a substantial challenge for tax compliance, it should not be interpreted primarily as a data quality issue. Instead, it reflects a structural characteristic of contemporary software development organizations, where teams are increasingly distributed, virtual, or composed of members affiliated with multiple subsidiaries. These organizational arrangements are precisely those that are most relevant from a taxation perspective. Consequently, statistical estimators or heuristics to infer missing jurisdictions would introduce systematic bias and undermine audit traceability. From a tax compliance standpoint, such cases require explicit investigation rather than approximation.

The discussion on ownership revealed a more fundamental conceptual tension. Although we defined ownership in the viewpoint as responsibility and accountability for a software component, the term ownership itself triggered strong legal associations among tax experts. In taxation, ownership is typically understood in terms of legal or economic ownership, which serves as the starting point for functional and risk analyses. Despite our explicit clarification, the architectural notion of ownership was repeatedly interpreted through this legal lens, leading to discussions that extend beyond the scope of this paper. This mismatch suggests that our definition of ownership is architecturally sound but legally underspecified. In particular, accountability in software engineering does not necessarily align with legal accountability in transfer pricing contexts. Addressing this gap requires interdisciplinary research that bridges software architecture concepts with legal and economic notions of ownership.

Another important insight concerns the quantitative representation of dependencies. Several experts interpreted weighted edges as implying that all components are normalized and equally relevant, which contradicts established taxation practice where components are not additive and value is highly context dependent. While this interpretation was unintended, it reveals a tension inherent in viewpoint design. Quantitative abstractions facilitate prioritization and scalability but risk being misread as value proxies. We maintain that weighted dependencies are useful for identifying areas of potential audit risk, as a higher number of dependencies implies greater uncertainty and therefore greater scrutiny. To mitigate misinterpretation, future iterations of the viewpoint may replace absolute counts with coarse-grained intervals, explicitly signaling that dependencies are indicators of relevance rather than measures of value. This observation aligns with recent empirical findings showing that ownership and contribution are unevenly distributed and that marginal owners tend to resolve issues more slowly in low-quality code bases \citep{borg2023u}.

Finally, the panel raised concerns regarding the temporal stability of software architectures. Our view represents a snapshot of the architecture at a single point in time, whereas tax audits typically cover extended fiscal periods. Empirical evidence on the rate and nature of architectural change, particularly in large-scale microservice systems, is scarce and likely context specific. However, given that the effort to generate the view is modest once data sources are established, regularly producing architectural snapshots appears feasible. We hypothesize that architectural evolution is often coupled with changes in ownership structures, either directly or indirectly. This suggests that longitudinal software architecture descriptions may be particularly valuable for tax compliance and warrants further investigation.

\section{Limitations}
\label{sec:limitations}

In this section, we critically discuss the limitations of our study and their impact on our findings. The limitations described below primarily affect the evaluation scope and the downstream applicability of our results, rather than the core contribution of this work, which is to investigate whether software architecture descriptions can make tax-relevant cross-border reuse structures visible to tax authorities.

\subsection{Anonymization}

The strict anonymization of the case company was a methodological prerequisite for conducting this study and for gaining access to highly sensitive architectural and organizational data. As a consequence, we had to remove context and potentially relevant information, including the specific business model, the revenue stream, and the organizational structure of our case company. Not being able to provide this information limits the generalizability of our evaluation. However, we believe that we provided sufficient context and architectural detail for the panel to deliver solid and high-level feedback suitable for the early stage of research addressed in this study.

To maintain strict anonymity, no internal representative of the case company participated in the expert panel. Instead, the author team provided the architectural stimulus based on data extracted from internal systems and prior collaboration with the company, without interpreting or evaluating taxation-related aspects on behalf of the organization. In this early stage of developing and evaluating a software architecture description, we consider this approach acceptable. In future work that addresses subsequent questions such as valuation of software components or transfer pricing, the involvement of internal stakeholders will, however, become crucial.

We further acknowledge that anonymization substantially influenced the quality and depth of the evaluation. In particular, we intentionally removed the economic and managerial context of the software architecture. Since software engineering does not occur in isolation from economic decision-making, this context plays a central role in taxation. The absence of such information led to several open questions during the evaluation. We therefore consider this a major limitation with respect to downstream questions of valuation and transfer pricing. At the same time, this limitation does not invalidate our findings regarding the ability of software architecture descriptions to surface tax-relevant cross-border reuse structures.

\subsection{Data Source for Software Architecture}

For creating the software architecture description, we relied on historical information extracted from an internal system that continuously collects the software product’s component structure, including microservices, from the code repositories. We consider this system a reliable and trustworthy data source, as it is also used internally for other operational purposes. The internal team responsible for maintaining the system supported us in interpreting the data and avoiding misinterpretations.

The date of extraction was randomly selected from the first two quarters of 2023. While microservice architectures may evolve continuously, tax audits and transfer pricing assessments typically reason over fiscal periods rather than instantaneous system states. We therefore consider a snapshot-based representation of the software architecture to be fit for the purpose of our evaluation, even though it cannot capture architectural changes over time.

\subsection{Panel Composition}

Both representatives of the tax authorities in our panel come from the same country. Although taxation standards relevant to transfer pricing are internationally harmonized and our panel consists of experts advising on international taxation, our study may not fully capture country-specific nuances in the interpretation and application of these standards. Consequently, our findings primarily apply to the fundamentals of describing software architectures for tax authorities.

However, since the OECD transfer pricing guidelines aim to establish a common framework for assessing cross-border transactions, we consider the panel appropriate for evaluating the general suitability and interpretability of software architecture descriptions for taxation purposes. Future research is required to investigate country-specific specializations and jurisdictional nuances in greater detail.

\section{Conclusion}
\label{sec:conclusion}

In this study, we investigated to what extent a deliberately minimal software architecture description can make the implicit cross-border licensing structure of globally distributed software systems visible and interpretable for tax experts. To this end, we defined a deliberately minimal architecture viewpoint framing tax-related concerns, constructed a view in accordance with it for a large-scale industrial microservice architecture, and evaluated the resulting software architecture description in a judgment study with experienced tax professionals. To our knowledge, this is the first empirical evaluation of a software architecture description explicitly constructed to support tax-related reasoning about cross-border software reuse.

The evaluation showed that the proposed software architecture description can make structurally relevant cross-border reuse relationships visible and provides a credible, evidence-based starting point for discussion with tax experts. In particular, the experts considered the software architecture description useful for surfacing potentially relevant parties, dependencies, and jurisdictional relationships that would otherwise remain implicit. At the same time, the evaluation also showed that the interpretability of such a description is limited in important ways. The experts consistently identified mismatches between software-engineering and legal notions of ownership, unclear jurisdictional assignments for distributed teams, and the risk of misinterpreting quantitative dependency counts as indicators of economic value.

Taken together, these findings indicate that a deliberately minimal software architecture description can make implicit cross-border licensing structures visible and partly interpretable for tax experts, while also revealing clear limits where software architecture abstractions alone are insufficient for tax-related reasoning. In this sense, the contribution of this work is not a complete solution for tax compliance, but an empirical characterization of the extent to which software architecture descriptions can support tax-related interpretation and where they systematically remain limited.

Although our evaluation focused on a microservice-based architecture, the challenges identified are not specific to this architectural style. Extending the approach to other architectural paradigms will require a more explicit and robust definition of software components, a challenge long recognized in software engineering due to the absence of a shared and context-independent understanding of what constitutes a software component (e.g.,~\citep{Broy1998}). More broadly, bridging the gap between software-engineering notions of ownership and the legal and economic notions used in taxation emerged as a key interdisciplinary challenge that warrants further research.

Finally, this work also highlights a broader methodological tension for empirical software engineering research on sensitive industrial and regulatory topics: strict anonymity and confidentiality may be prerequisites for access, candor, and validity, but they constrain disclosure and reproducibility in the traditional sense. We nevertheless argue that our findings provide transferable insight into both the potential and the limits of using software architecture descriptions to make cross-border reuse visible and interpretable for tax experts.

\authorcontributions{Michael~Dorner: Conceptualization, Data curation, Formal analysis, Investigation, Methodology, Software, Visualization, Writing – original draft;
Oliver~Treidler: Conceptualization, Formal analysis, Investigation, Writing – review and editing;
Tom-Eric~Kunz: Conceptualization, Investigation, Validation, Writing – review and editing;
Ehsan~Zabadast: Data curation, Software, Writing – review and editing;
Daniel~Mendez: Funding acquisition, Methodology, Writing – review and editing;
Darja~Šmite: Visualization, Writing – original draft;
Maximilian~Capraro: Conceptualization, Validation, Visualization, Writing – review and editing;
Krzysztof~Wnuk: Writing – review and editing;
All authors have read and agreed to the published version of the manuscript.}

\funding{This work was supported by the Knowledge Foundation (KK-stiftelsen) through the SERT~Project (Research Profile Grant 2018/010) at Blekinge Institute of Technology.
}

\informedconsent{Informed consent was obtained from all subjects involved in the study.}

\dataavailability{%
Due to the sensitivity of the industrial architectural data and the confidentiality of the expert interviews, the underlying raw data used in this study cannot be shared. To support transparency and enable methodological scrutiny, the paper describes the software architecture viewpoint, the procedure used to construct the evaluation stimulus, and the judgment study design in detail. The interview guidelines are publicly available \citep{Dorner2026}. No raw architectural data, company-identifying information, or interview transcripts are shared.
} 

\useofartificialintelligence{The authors used an AI-based language model to support language editing and improve grammar and clarity. The AI tool was used solely for stylistic refinement; all scientific content, analysis, and conclusions were developed and verified by the authors.}

\acknowledgments{We would like to thank our industry partner and the panel of tax experts for their great support, which made this research possible.}

\conflictsofinterest{The authors declare no conflicts of interest.}

\bibliography{references.bib}

\end{document}